\documentclass[prx,twocolumn,superscriptaddress,floatfix,nopacs]{revtex4}
\usepackage{graphicx,amsfonts,amssymb,amsmath,hyperref,enumerate}

\newif\ifhyper
\hypertrue
\ifhyper
\hypersetup{
  citecolor = {green},
  colorlinks = {true}, 
  urlcolor = {blue} 
}
\fi

\newcommand{\beq}{\begin{equation}}
\newcommand{\eeq}{\end{equation}}
\newcommand{\beqa}{\begin{eqnarray}}
\newcommand{\eeqa}{\end{eqnarray}}
\newcommand{\ket} [1] {\vert #1 \rangle}

\def\ket#1{\vert#1\rangle}

\def\Longarrow{\protect\@lra}
\def\@lra{\relbar\joinrel\relbar\joinrel\relbar\joinrel%
     \relbar\joinrel\rightarrow}

\begin{document}

\title{Benchmarking global $SU(2)$ symmetry in 2d tensor network algorithms}
\author{Philipp Schmoll}
\affiliation{Institute of Physics, Johannes Gutenberg University, 55099 Mainz, Germany}
\author{Rom\'an Or\'us}
\affiliation{Donostia International Physics Center, Paseo Manuel de Lardizabal 4, E-20018 San Sebasti\'an, Spain}
\affiliation{Ikerbasque Foundation for Science, Maria Diaz de Haro 3, E-48013 Bilbao, Spain}
\affiliation{Multiverse Computing, Paseo de Miram\'on 170, E-20014 San Sebasti\'an, Spain}

\begin{abstract}

We implement and benchmark tensor network algorithms with $SU(2)$ symmetry for systems in two spatial dimensions and in the thermodynamic limit. Specifically, we implement $SU(2)$-invariant versions of the infinite Projected Entangled Pair States (iPEPS) and infinite Projected Entangled Simplex States (iPESS) methods. Our implementation of $SU(2)$ symmetry follows the formalism based on fusion trees from [P. Schmoll, S. Singh, M. Rizzi, R. Or\'us, Annals of Physics \textbf{419}, 168232]. In order to assess the utility of implementing $SU(2)$ symmetry, the algorithms are benchmarked for three models with different local spin: the spin-$1$ bilinear-biquadratic model on the square lattice, and the Kagome Heisenberg antiferromagnets (KHAF) for spin-$1/2$ and spin-$2$. We observe that the implementation of $SU(2)$ symmetry provides better energies in general than non-symmetric simulations, with smooth scalings with respect to the number of parameters in the ansatz, and with the actual improvement depending on the specifics of the model. In particular, for the spin-$2$ KHAF model, our $SU(2)$ simulations are compatible with a quantum spin liquid ground state. 

\end{abstract}

\maketitle

{\color{blue} {\it Introduction.-}} Tensor networks~\cite{TNrevs} (TN) are mathematical objects tailored to describe highly-correlated structures in an efficient way. In condensed matter physics they can describe low-energy states of quantum matter. The success of TN methods has been particularly impressive for one-dimensional (1d) systems \cite{DMRG}, with the Matrix Product State (MPS) as the driving force. On top of that, many applications of TN methods have also been developed to tackle strongly correlated systems in two spatial dimensions (2d). Here, Projected Entangled Pair States (PEPS)~\cite{PEPS} are widely used, and the infinite-PEPS (iPEPS) algorithm~\cite{iPEPS} is nowadays a standard tool for simulations in the thermodynamic limit. Alternative methods such as the infinite Projected Entangled Simplex States (iPESS)~\cite{iPESS} have been applied with success to the Kagome lattice~\cite{iPESS_various}. 

An important problem in tensor networks, especially in 2d algorithms like iPEPS and iPESS, is how to deal with global non-abelian symmetries, $SU(2)$ being a common example. For instance, numerical simulations of the \mbox{spin-$1/2$} Kagome Heisenberg Antiferromagnet (KHAF) seem to indicate~\cite{KHAF} that its ground state is a quantum spin liquid and therefore an $SU(2)$ singlet. One would therefore expect, a priori, that the study of such a ground state with a TN algorithm would benefit from the explicit preservation of $SU(2)$ symmetry. While this has been done already using an $SU(2)$-invariant implementation of DMRG~\cite{SU(2)DMRG}, the generalization to true 2d TN algorithms has been achieved only in a very few cases~\cite{SU(2)PEPS}.

In this paper we implement $SU(2)$ symmetry in iPEPS and iPESS algorithms using the formalism from Ref.~\cite{OurSU(2)}, which is based on fusion trees. We benchmark our implementation by computing ground state properties of three 2d models: the spin-$1$ bilinear-biquadratic model on the square lattice, and the spin-$1/2$ and spin-$2$ KHAF. We observe that the implementation of $SU(2)$ in the 2d simulations in general allows to produce lower energies than the ones obtained using non-symmetric TN algorithms. However, since $SU(2)$-invariant tensors are highly constrained, we find that the actual improvement depends a lot on the specifics of the model. In particular, for the spin-$2$ KHAF model, the $SU(2)$ simulations produce a ground state structure compatible with that of a quantum spin liquid to the best of our computational power.

\vspace{5pt}

{\color{blue} {\it Methods.-}} We implemented $SU(2)$-invariant versions of iPEPS and iPESS algorithms. We refer the interested reader to Ref.~\cite{iPEPS} for details about iPEPS, and to Ref.~\cite{iPESS} about iPESS. Let us just mention that, in this paper, we employ the so-called \emph{simple update}~\cite{su}, which provides an efficient tensor update for an imaginary-time evolution algorithm, also when combined with $SU(2)$ symmetry. The accuracy of our calculations could always be improved by more precise tensor optimization schemes~\cite{iPEPS, ffu, vu}, but at the cost of extra computational expense. Expectation values in all cases are approximated using well-known Corner Transfer Matrix (CTM) techniques~\cite{CTM}, which can also be easily adapted to deal with $SU(2)$. 

\begin{figure}
	\centering
	\includegraphics[width = 1\columnwidth]{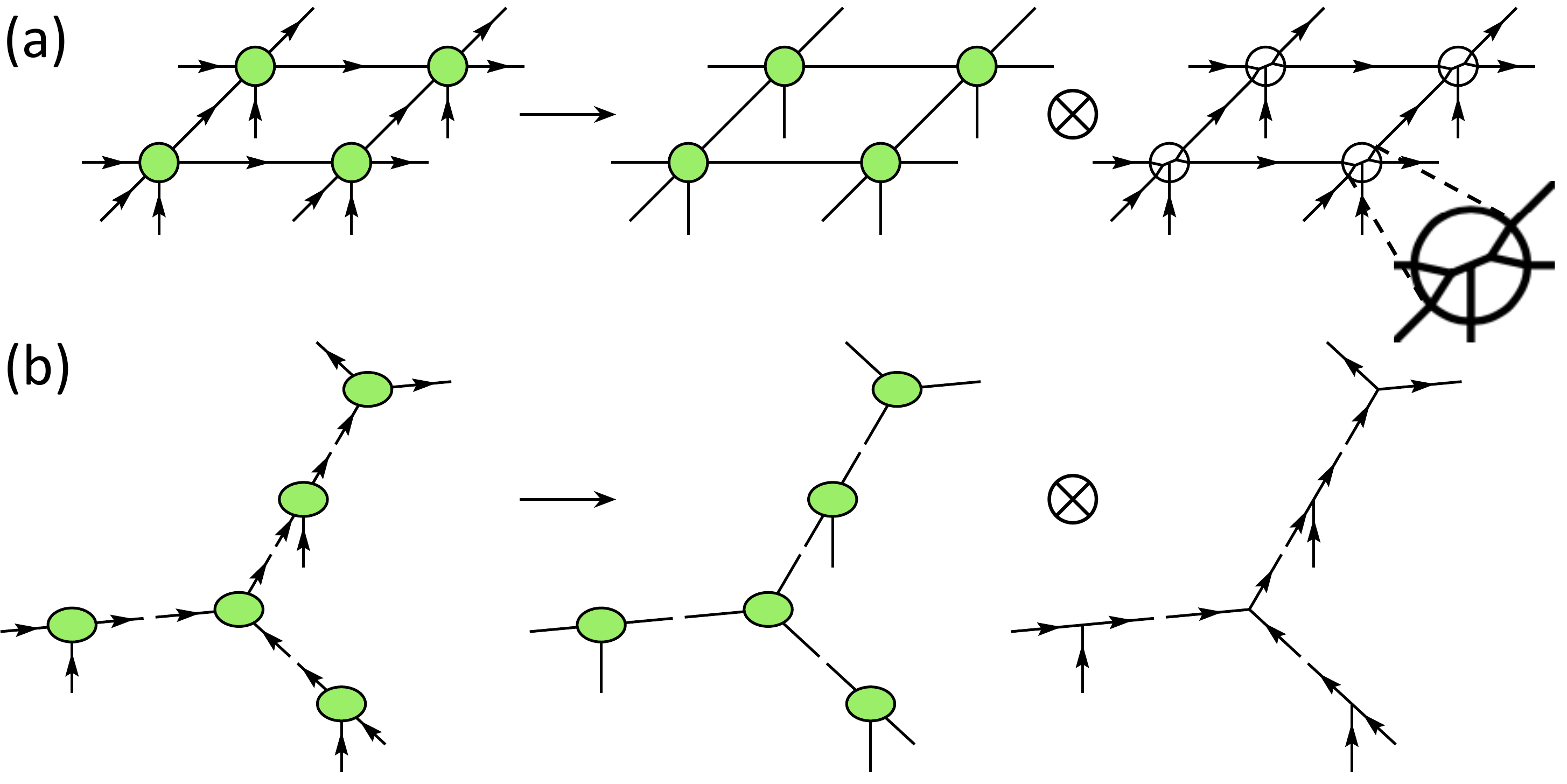}
	\caption{(Color online) (a) Decomposition of an $SU(2)$-symmetric $2 \times 2$ iPEPS unit cell on a square lattice, in terms of degeneracy tensors and a network of fusion trees. The fusion tree structure of every tensor is shown on the right-hand side, where arrows correspond to incoming/outgoing indices. Every three-index node in the fusion trees is an intertwiner of $SU(2)$, i.e., a tensor of Clebsch-Gordan coefficients. (b) Decomposition of an $SU(2)$-symmetric iPESS unit cell on the honeycomb lattice, which is used to simulate its dual Kagome lattice.}
	\label{Fig1}
\end{figure}

Concerning $SU(2)$ itself, we work here with the implementation from Ref.~\cite{OurSU(2)}. Under the action of the group, both the physical and the virtual vector spaces of the tensors can be described in terms of basis states $\ket{j,t_j,m_j}$. Here $j$ are the irreducible representations of $SU(2)$, i.e. the spin quantum numbers $j = 0,1/2,1,\hdots$, $t_j$ labels the multiple copies of irrep $j$, and $m_j$ labels the states within the vector space of spin $j$ ($m_j = \lbrack -j,\hdots,+j \rbrack$). Due to the action of the group, a symmetric tensor generally consists of various symmetric blocks, each of which factorizes into a degeneracy part and a structural part according to the Wigner-Eckard theorem. The degeneracy part contains the remaining variational parameters, the structural part is completely determined by the underlying symmetry and describes the coupling of the spins on the tensor indices in terms of Clebsch-Gordan coefficients. In our implementation the structural part of the symmetric tensors is codified in the form of fusion trees decorated by quantum numbers. They are a memory-efficient analytic representation of the group structure, so that no actual Clebsch-Gordan coefficients for the structural tensors have to be stored. In Fig.~\ref{Fig1} we show the decomposition of a network of iPEPS and iPESS tensors in terms of degeneracy parts and structural parts. This implementation allows for clean, accurate calculations, which is of particular importance when dealing with 2d TN algorithms. Concerning notation, in the following we call $D_\text{eff}$ the effective bond dimension of the PEPS or PESS, i.e., $D_\text{eff} = D$ for non-symmetric TNs (with $D$ the usual bond dimension) and $D_\text{eff} = \sum_{j_i} t_{j_i} \times |m_{j_i}|$ for symmetric ones (with $t_{j_i}$ the degeneracy of symmetry sector $j_i$ and $|m_{j_i}| = 2j_i +1$ for an index $i \equiv (j_i, t_j, m_{j_i})$ \cite{OurSU(2)}). The symmetric bond dimension is $D_\text{sym} = \sum_{j_i} t_{j_i}$ and $N$ is the number of variational parameters in the ansatz.

\vspace{5pt}

{\color{blue} {\it Results.-}} Let us now discuss the performance of $SU(2)$-invariant iPEPS and iPESS. We focus first on the \mbox{spin-$1$} bilinear-biquadratic (BLBQ) model on the square lattice. Its Hamiltonian is given by
\begin{align}
	H = \sum_{\langle i,j \rangle} \left( \cos(\theta) \, \left( \boldsymbol S_i \cdot \boldsymbol S_j \right) + \sin(\theta) \, \left( \boldsymbol S_i \cdot \boldsymbol S_j \right)^2 \right),
\end{align}
where $\langle i,j \rangle$ are nearest-neighbour interactions, $\boldsymbol S_i$ is the vector of spin-$1$ matrices, and $\theta$ tunes the relative coupling strength of the bilinear and biquadratic terms. The phase diagram of this model has already been computed previously with iPEPS, both without symmetries but also including $U(1)$ symmetry \cite{BBCorboz}. Here, we tune the coupling parameter to $\theta = 0.21\pi$. At this angle the magnetization of the ground state vanishes, so that the system is believed to be in the middle of an $SU(2)$-symmetric Haldane phase existing for $\theta \in \lbrack 0.189 , 0.217 \rbrack$. This phase is reminiscent of coupled spin-$1$ chains~\cite{BBCorboz}. Thus, the point $\theta = 0.21\pi$ is a non-trivial benchmark for a 2d $SU(2)$-invariant gapped phase. In our simulation, we choose to work with the simple update and a $2 \times 2$ unit cell. 
\begin{figure}
	\centering
	\includegraphics[width = \columnwidth]{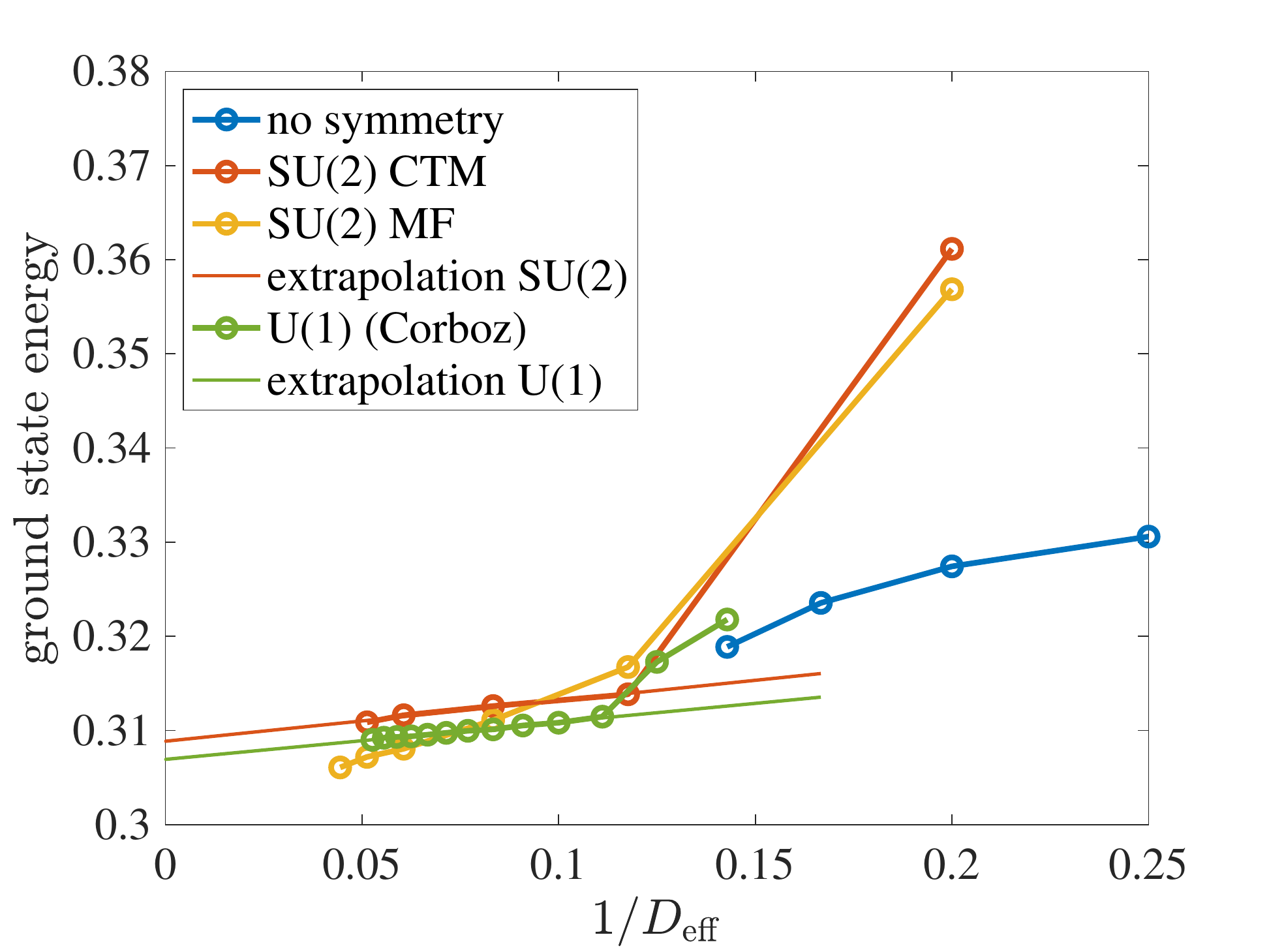}
	\caption{Ground state energy of the spin-$1$ BLBQ model on the square lattice at $\theta = 0.21\pi$, as a function of $1/D_\text{eff}$. $U(1)$ results are from Ref.~\cite{BBCorboz} and replotted with permission.}
	\label{Fig2}
\end{figure}
\begin{figure}
	\centering
	\includegraphics[width = \columnwidth]{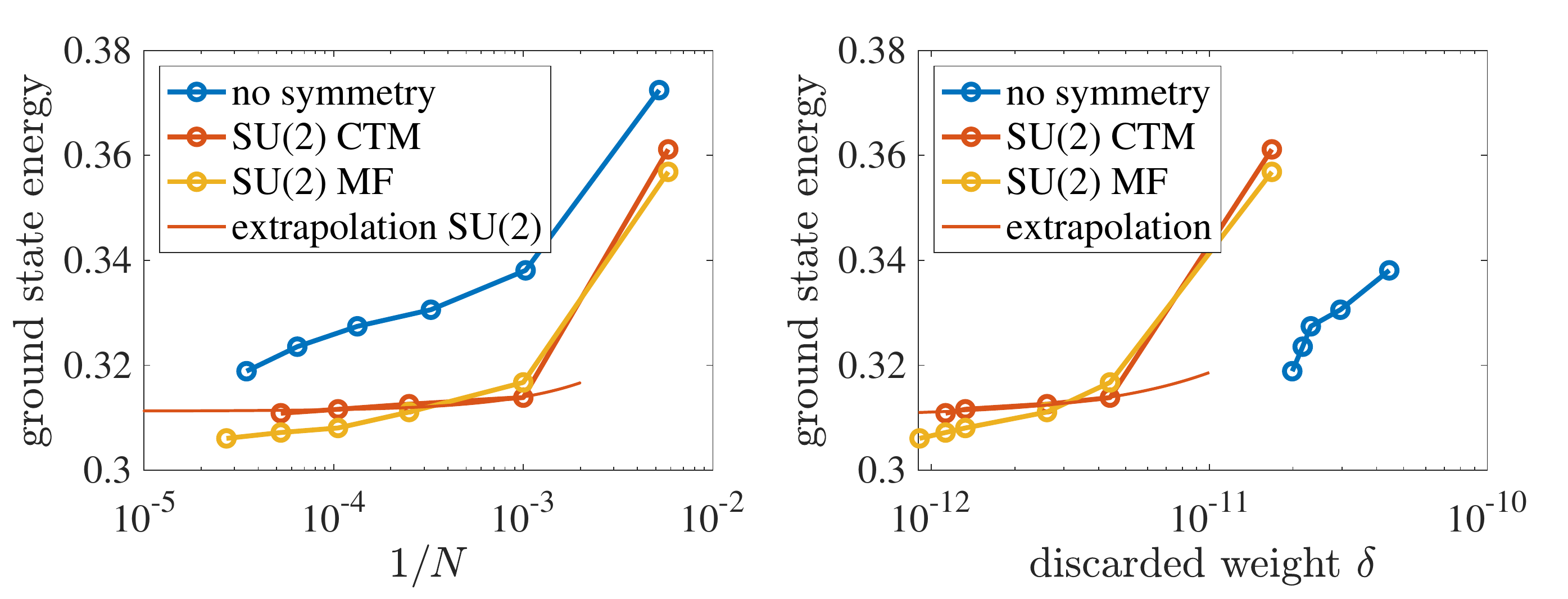}
	\caption{(Color online) Ground state energy of the spin-$1$ BLBQ model on the square lattice at $\theta = 0.21\pi$, as a function of $1/N$ (left) and as a function of the discarded weight $\delta$ (right).}
	\label{Fig3}
\end{figure}
The ground state energy of the system is shown in Fig.~\ref{Fig2} as a function $1/D_\text{eff}$, in Fig.~\ref{Fig3} as a function of $1/N$ and as a function of the discarded weight in the truncations $\delta$~\cite{discarded}. The plots show the performance for $SU(2)$-iPEPS as well as for iPEPS with no symmetry, and we also compare with the results from Ref.~\cite{BBCorboz} using $U(1)$-iPEPS in Fig.~\ref{Fig2} (results replotted with permission). In the figures, for the $SU(2)$ simulations we include results obtained by using a CTM environment to compute expectation values, as well as using a mean-field (MF) environment estimation. This last option does not provide variational energies, but allows us to see the overall tendency for large bond dimension (for which the calculations using $SU(2)$-CTM algorithms are computationally costly). We see that the extrapolation $1/N \rightarrow 0$ is better behaved than the one for $1/D_\text{eff} \rightarrow 0$, and is actually comparable for $SU(2)$ to the extrapolation in the discarded weight. In this last extrapolation one can also clearly see that the non-symmetric simulation is far from being converged. Our extrapolated data for the ground state energy $e_0$ is $e_0(1/D_\text{eff} \rightarrow 0) = 0.309 \pm 0.003$, $e_0(1/N \rightarrow 0) = 0.311 \pm 0.004$, and $e_0(\delta \rightarrow 0) = 0.310 \pm 0.002$. 

We notice from our plots that the simulations without symmetry yield the lowest ground state energy for small bond dimensions and the data points with $SU(2)$ symmetry are considerably higher than those with lower or no symmetry. We take this as a first indication that the $SU(2)$-symmetric ansatz in 2d may sometimes be too restrictive, which is especially true for small bond dimensions. However, for large bond dimension the situation is the opposite, and the $SU(2)$ simulation produces lower energies. It is interesting, though, that the $SU(2)$ numbers computed by CTM (which are variational, since the CTM bond dimension is converged \cite{CTM}) tend to be always slightly above those obtained with an $U(1)$-symmetric ansatz. We will comment on the restrictiveness and expressiveness of our simulations in the next paragraph.
\begin{figure}
	\centering
	\includegraphics[width = 0.6\columnwidth]{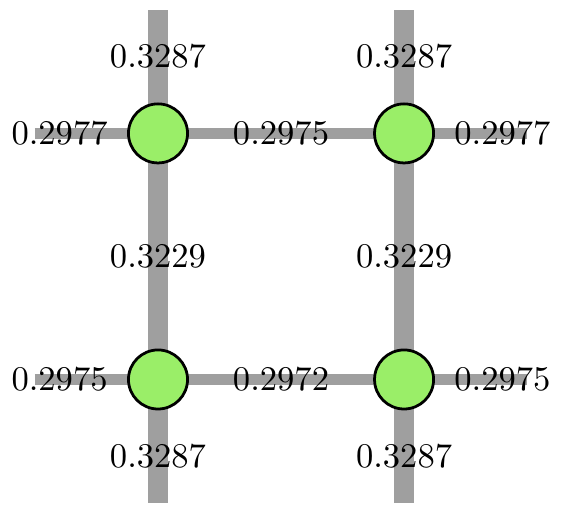}
	\caption{(Color online) Ground state energy per link in the iPEPS unit cell. The structure is compatible with vertical Haldane chains coupled in the horizontal direction. The differences in the fourth relevant digit between the upper and lower horizontal link energies is due to truncation effects.}
	\label{Fig4}
\end{figure}
For the record, the obtained extrapolated energy with $U(1)$ symmetry in Fig.~\ref{Fig2} is $e_0(1/D_\text{eff} \rightarrow 0) = 0.307 \pm 0.001$, and therefore very close to the $SU(2)$ number. Finally, in order to understand better the nature of the $SU(2)$-invariant ground state that we obtain, we also plot its energy on each link of the iPEPS unit cell in Fig.~\ref{Fig4}. The ground state cultivates different energies in $x$- and $y$-directions, thus breaking lattice rotation symmetry. This is however compatible with vertical coupled Haldane 1d chains, in accordance with the results from Ref.~\cite{BBCorboz}. This difference in bond energies is caused by converging to half-integer spin representations on the vertical bonds, and integer ones on the horizontal bonds, leading to different effective bond dimensions.\\
\begin{figure}
	\centering
	\includegraphics[width = 0.7\columnwidth]{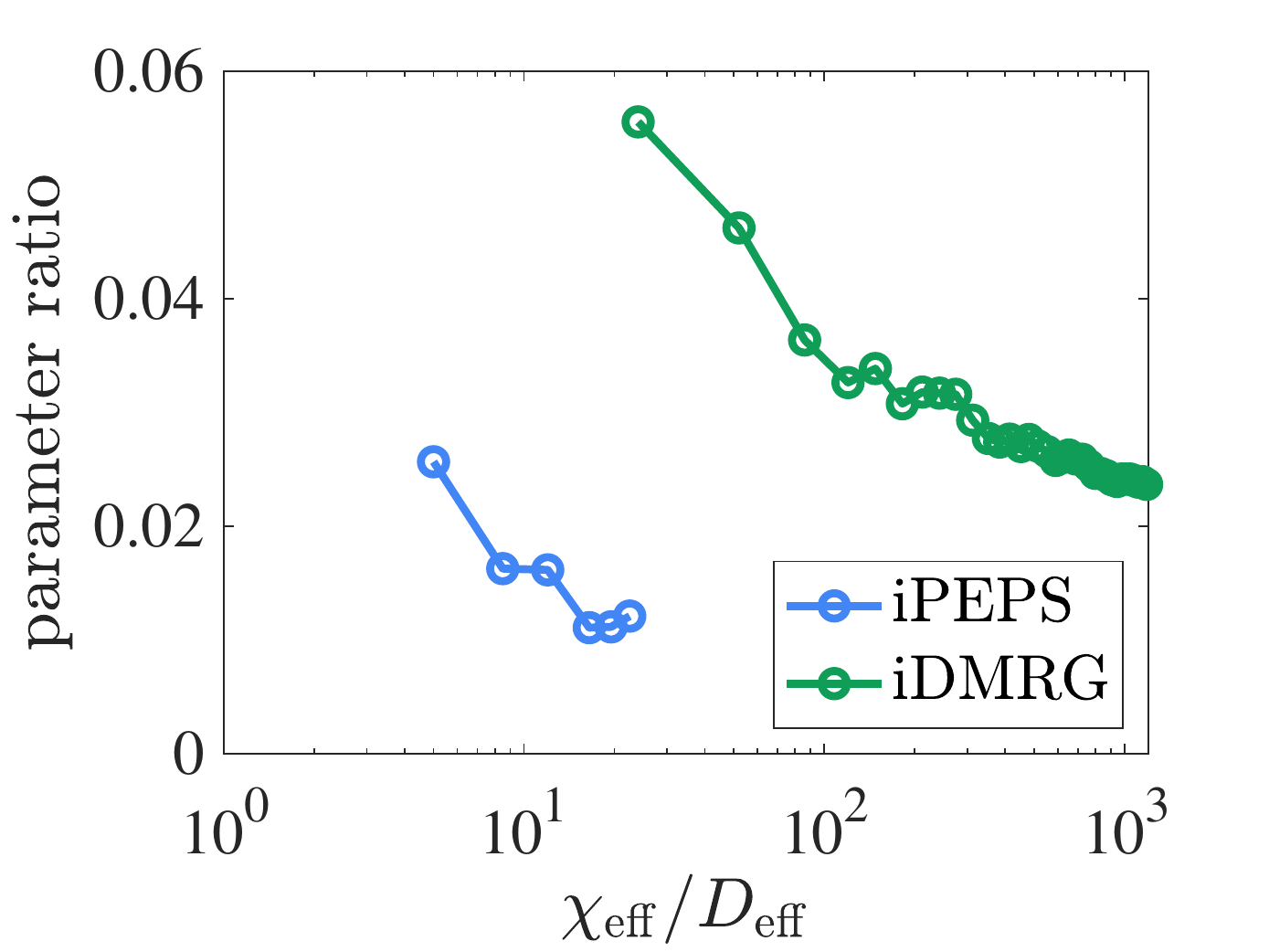}
	\caption{(Color online) Ratio of ansatz variational parameters between $SU(2)$-symmetric and non-symmetric simulations, for iPEPS and iMPS, as a function of their respective bond dimensions $D_\text{eff}$ and $\chi_\text{eff}$.}
	\label{Fig5}
\end{figure}
These findings point towards an interesting fact: $SU(2)$ symmetry in 2d, even if generically useful, can be highly restrictive in some cases. The variational space is highly constrained, and in some situations this could be too limited to find a good approximation to the ground state with ``simple" tensor updates. In order to get an idea of the effect of the symmetry on the size of the variational space we evaluated the ratio between remaining variational parameters in the $SU(2)$-iPEPS and the number of variational parameters in the corresponding unconstrained TN for different bond dimensions. This is shown in Fig.~\ref{Fig5}, alongside with the same information for an $SU(2)$-symmetric infinite MPS simulation of a critical spin-$1/2$ ladder system~\cite{ladderSU(2)}. The comparison between both cases allows us to understand better the effect of dimensionality in the reduction of variational parameters in a $SU(2)$-invariant TN ansatz. What we conclude from the plot is that the $SU(2)$-invariant ansatz becomes very restrictive with the bond dimension, as expected, but at a much faster rate in 2d than in 1d. In other words, $SU(2)$ in 2d restricts the variational space faster than in 1d. A priori, this could be good news, since the number of parameters to optimize is much more drastically reduced in 2d than in 1d. However, this needs to be taken into account with care when assessing symmetric TN simulations since the optimization space may actually be too constrained in some cases for finding low variational energies. The systematically higher $SU(2)$ energies could also hint at the fact, that the ground state weakly breaks the symmetry, albeit predicted to be symmetric. In this scenario, a manifestly $SU(2)$-invariant TN ansatz is expected to yield higher energies than an ansatz with a lower symmetry.\\

The next model that we considered was the spin-$1/2$ KHAF. The Hamiltonian is given by
\begin{align}
	H = \sum_{\langle i,j \rangle} \boldsymbol S_i \cdot \boldsymbol S_j\ ,
\end{align}
where $\langle i,j \rangle$ denotes nearest-neighbour interactions between sites of the Kagome lattice, and $\boldsymbol S_i$ is the \mbox{spin-$1/2$} (vector) operator at site $i$. The Kagome lattice exhibits corner-sharing triangles resulting in huge quantum fluctuations around the ground state due to strong geometric frustration, with many states very close in energy and competing to be the true ground state. This makes the simulation of the model very challenging.
\begin{figure}
	\centering
	\begin{minipage}{0.85\columnwidth}
		\includegraphics[width = \textwidth]{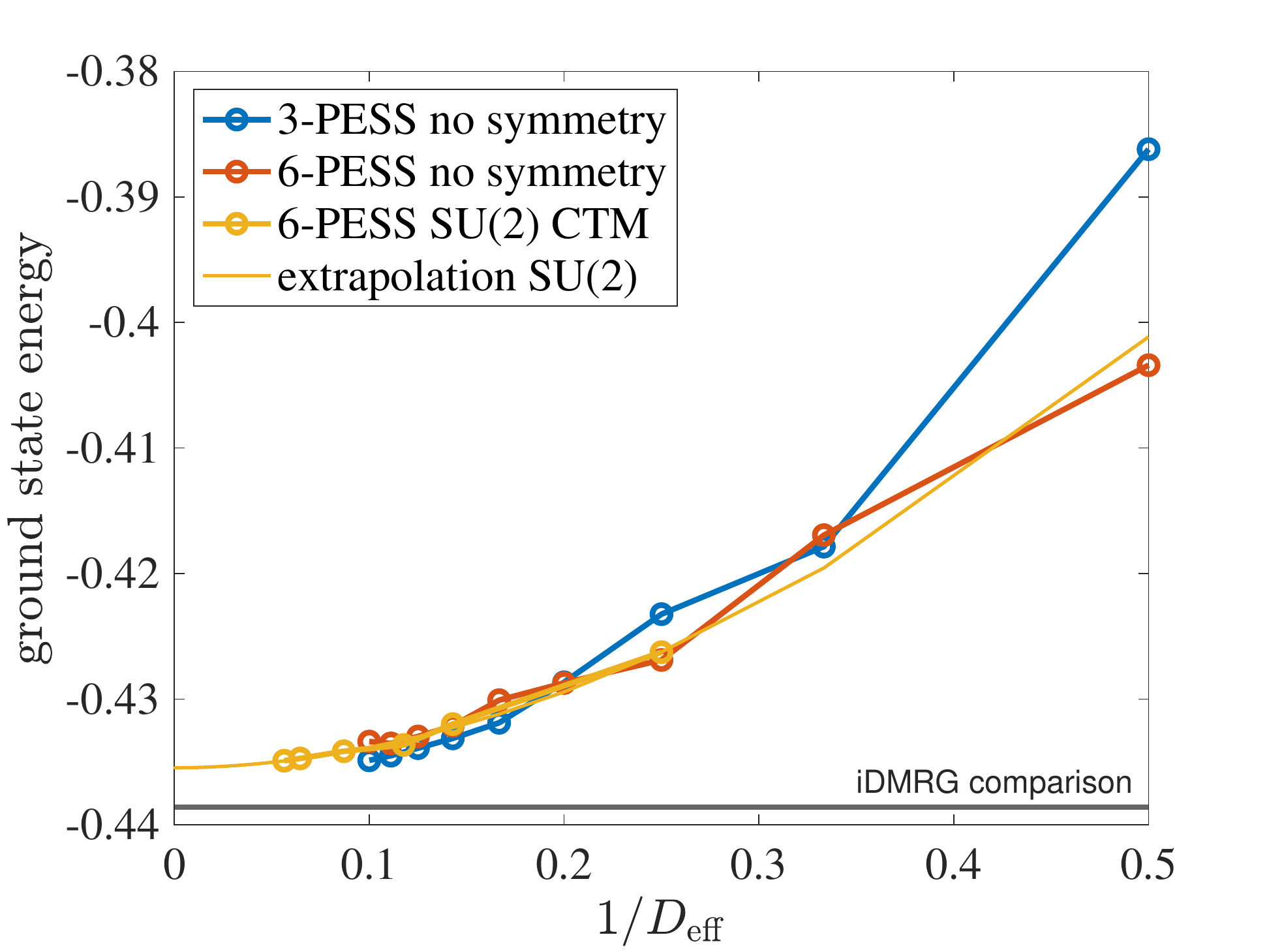}
	\end{minipage}
	\begin{minipage}{0.85\columnwidth}
		\includegraphics[width = \textwidth]{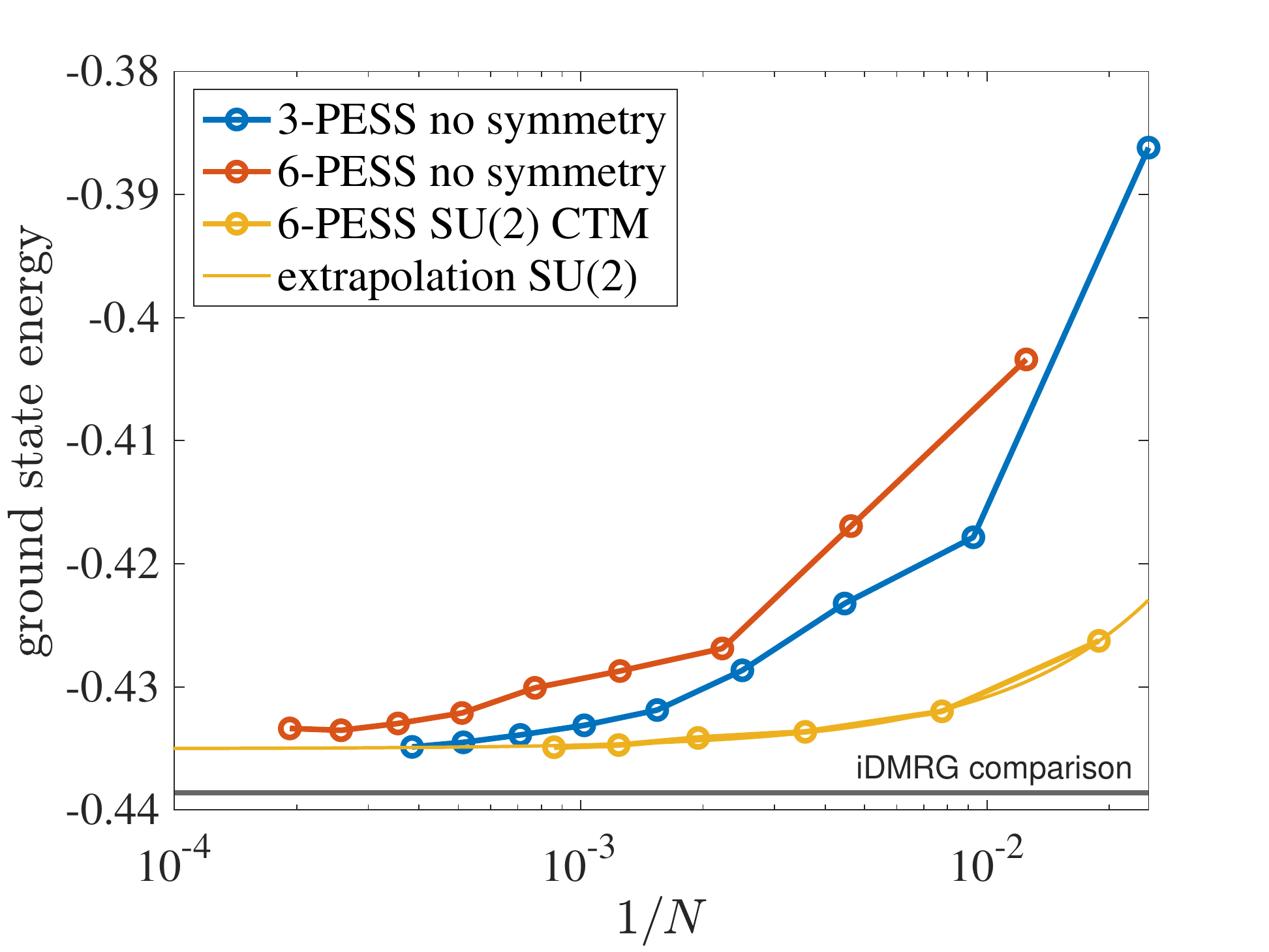}
	\end{minipage}
	\caption{(Color online) Ground state energy of the spin-$1/2$ KHAF, as a function of $1/D_\text{eff}$ and $1/N$, with the yellow line denoting the extrapolation to infinite bond dimension.}
	\label{Fig6}
\end{figure}
For the sake of this study, our goal here is \emph{not} to provide better ground-state numbers than those obtained by other simulations~\cite{KHAF}, but rather to benchmark the utility of $SU(2)$ symmetry in 2d TN algorithms, and in particular in iPESS. Previous results have shown that using three-site iPESS without symmetries produces reasonably good numbers for the ground state energy~\cite{iPESS}. For the $SU(2)$-symmetric simulations, however, we need to resort to the six-site unit cell in order to accommodate consistently the $SU(2)$ quantum numbers on all the indices of the symmetric TN ansatz. Since the physical sites carry spin-$1/2$, the geometry and the unit cell force us to use mixed spins (integer \emph{and} half-integer) on the bond indices of the iPESS.

\begin{figure}
	\centering
	\includegraphics[width = 1\columnwidth]{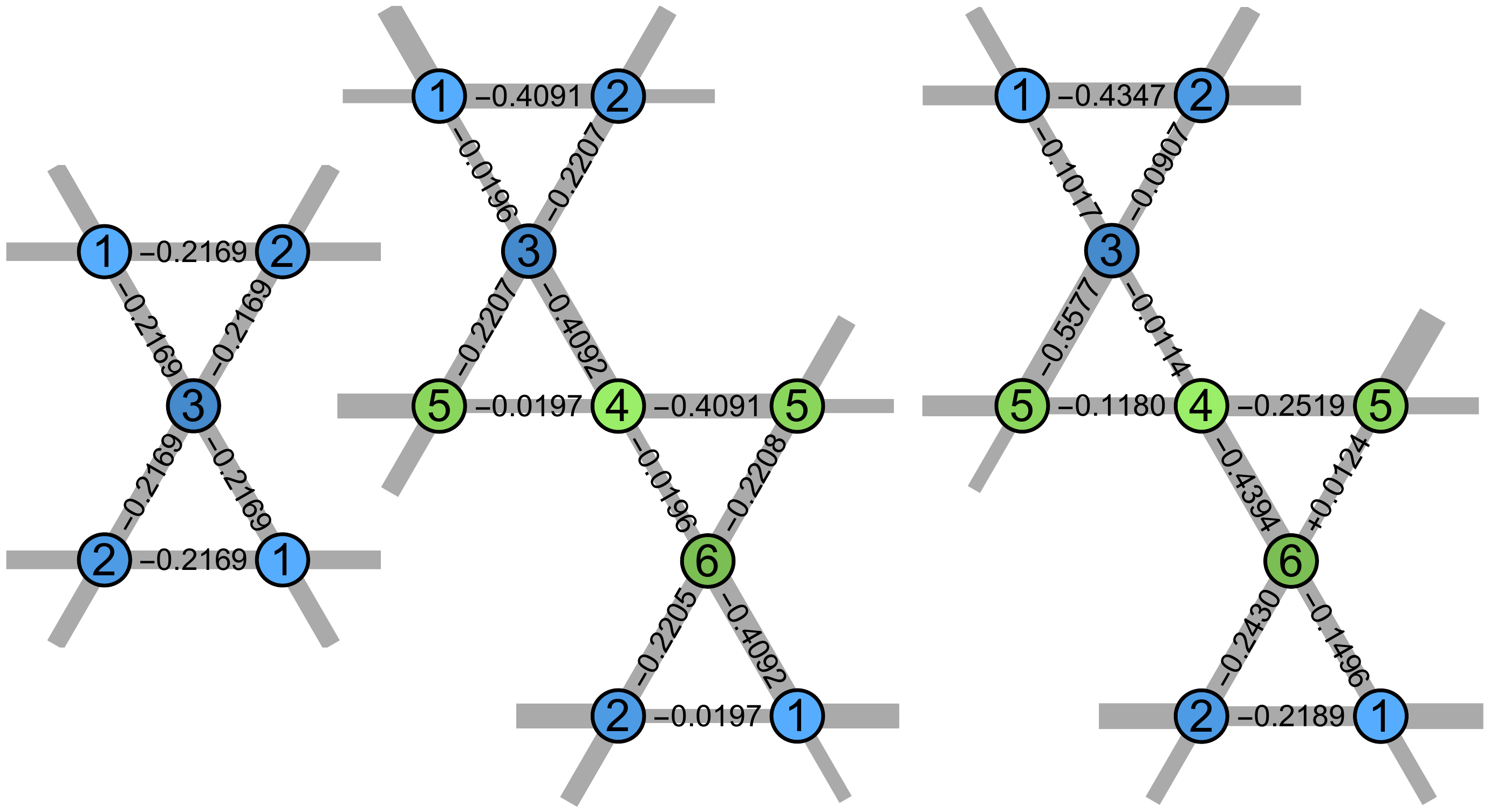}
	\caption{(Color online) Spin-spin correlation $\langle \boldsymbol S_i \boldsymbol S_j \rangle$ on each link of the unit cell for the non-symmetric 3-PESS, the non-symmetric 6-PESS and the $SU(2)$-invariant 6-PESS (from left to right).}
	\label{Fig7}
\end{figure}

\begin{figure}
	\centering
	\includegraphics[width = 1\columnwidth]{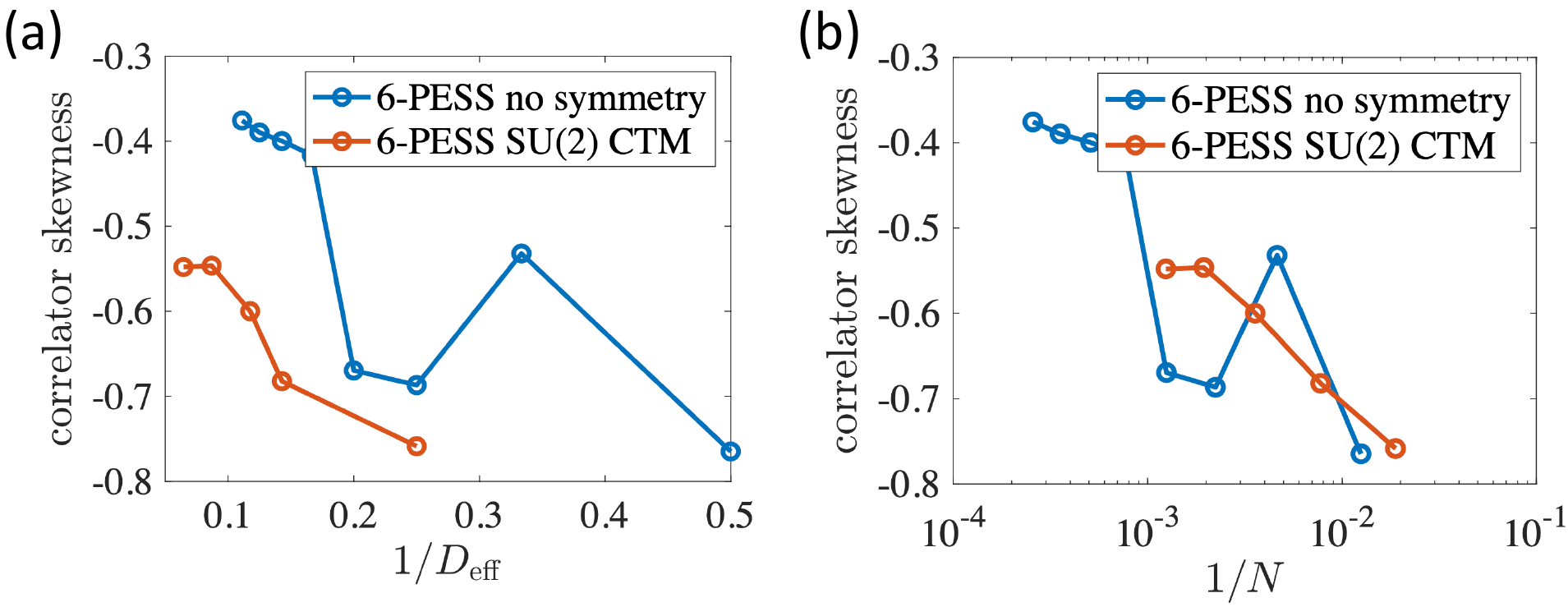}
	\caption{(Color online) Difference between the strongest and the weakest bond in the unit cell (correlator skewness) of a 6-PESS, for the spin-$1/2$ KHAF simulations, with no symmetry and with $SU(2)$ symmetry, as a function of (a) $1/D_\text{eff}$ and (b) $1/N$.}
	\label{Fig8}
\end{figure}

We computed the ground state energy of the model for a 3-site and a 6-site unit cell without symmetry, and for a 6-site unit cell with $SU(2)$ symmetry. The results are shown in Fig.~\ref{Fig6}. The symmetric results are compatible with those obtained without symmetries, with an algebraic convergence of the ground state energy as a function of $1/D_\text{eff}$ (see Fig.~\ref{Fig6}(a)), in turn reinforcing the observation that the ground state is a quantum spin liquid and therefore $SU(2)$-invariant. However, one can see again that the limit of infinite bond dimension is better achieved by the $SU(2)$-invariant simulations as a function of $1/N$ (see Fig.~\ref{Fig6}(b)), with extrapolated values $e_0(1/D_\text{eff} \rightarrow 0) = -0.435 \pm 0.004$ and $e_0(1/N \rightarrow 0) = -0.435 \pm 0.002$. The iDMRG comparison shows the current energetically lowest ground state energy~\cite{iDMRGLowestEnergy}. Let us mention that for all the iPESS simulations that we performed, the extrapolation in the discarded weight was not possible because the discarded weight was always too small. The spin-spin correlators for each link of the unit cell are shown in Fig.~\ref{Fig7}, for the non-symmetric 3-PESS, the non-symmetric 6-PESS and the $SU(2)$-invariant 6-PESS respectively.
While the 3-site unit cell produces a state that seems compatible with a quantum spin liquid, the 6-site unit cells seem to produce valence-bond crystal structures with strong and weak links, thus breaking invariance under translations and lattice rotations. We observe, in any case, that the valence bond crystal tends to melt when we increase the bond dimension of the iPESS ansatz, both for the non-symmetric and the $SU(2)$-invariant simulations, thus slowly recovering translation invariance (see Fig.~\ref{Fig8} for plots of the correlator skewness as $D_\text{eff}$ increases).
Interestingly, we also observe in the figure that the non-symmetric simulations tend to melt faster than the $SU(2)$-symmetric ones. The observed pattern of local correlations for the non-symmetric 6-site iPESS in Fig.~\ref{Fig7} is expected, since TN simulations tend to trade symmetry for injectivity of the target state. For the symmetric 6-site iPESS the correlators are even more skewed presumably due to non-uniform effective bond dimensions caused by the mixture of integer and half-integer representations. This effect is expected to vanish in the limit of large bond dimensions, and also explains the slower melting in Fig.~\ref{Fig8}.

\begin{figure}
	\centering
	\includegraphics[width = 1\columnwidth]{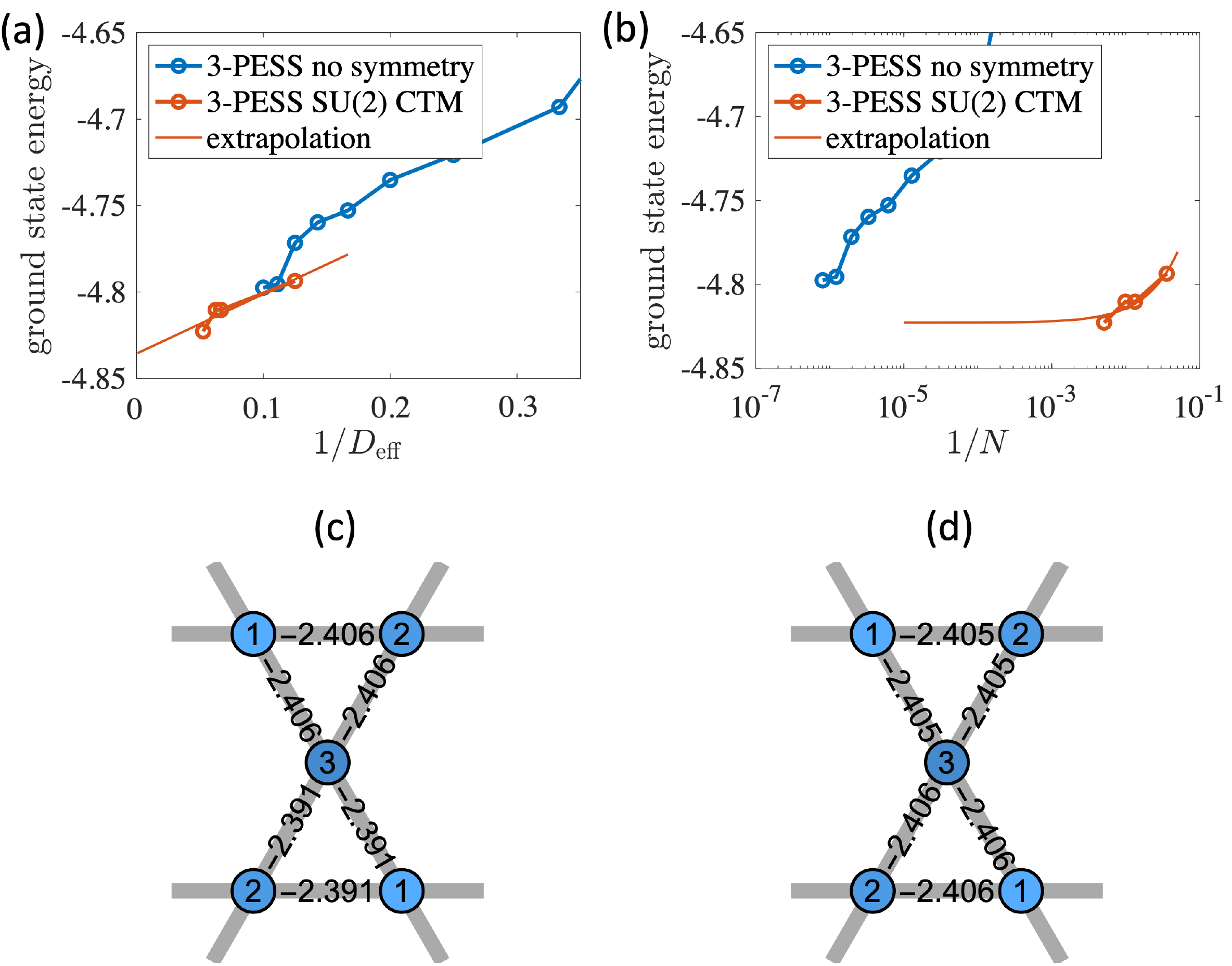}
	\caption{(Color online) Ground state energy of the spin-$2$ KHAF, as a function of (a) $1/D_\text{eff}$ and (b) $1/N$, with the red line denoting the extrapolation to infinite bond dimension. Notice that in this case, this extrapolation is just for completeness since the numbers still do not show convergence for the achievable bond dimensions due to the large local spin at every site. (c,d) Spin-spin correlation $\langle \boldsymbol S_i \boldsymbol S_j \rangle$ on each link of the unit cell for the non-symmetric and the $SU(2)$-invariant 3-PESS respectively.}
	\label{Fig9}
\end{figure}

Finally, we computed the ground state energy of the spin-$2$ KHAF for a 3-site unit cell without and with $SU(2)$ symmetry. Unlike in the spin-$1/2$ case, the fact that we have spin-$2$ in the physical indices allows us to use the 3-site unit cell (this, in fact, is true for all integer-spin Heisenberg models on the Kagome lattice). The results are shown in Fig.~\ref{Fig9}. This time, due to the large dimension of the physical spin at every site, we cannot reach values of $D_\text{sym}$ as large as for the spin-$1/2$ case. However, the effective bond dimension is larger for the spin-$2$ KHAF, as shown in Table~\ref{tab0}. We see in Figs.~\ref{Fig9}(a,b) that the $SU(2)$-invariant simulations can handle larger $D_\text{eff}$ and produce lower variational energy than the non-symmetric simulations. Moreover, both in the non-symmetric and symmetric cases we observe in Figs.~\ref{Fig9}(c,d) a structure of spin-spin correlators in the unit cell that seems compatible with that of a quantum spin liquid, which is also compatible with the algebraic behaviour of the ground state energy with $D_\text{eff}$ in Fig.~\ref{Fig9}(a). The energies are, however, difficult to extrapolate to infinite bond dimension, and hence we cannot be sure whether this is the true nature of the ground state. But we can claim that, to the best of our calculations, here the $SU(2)$-invariant iPESS with a 3-site unit cell produces the best variational energy for the ground state, which seems compatible with a quantum spin liquid. Moreover, we computed the expectation value of the chiral correlators $\boldsymbol S_i \cdot ( \boldsymbol S_j \times \boldsymbol S_k)$ on all triangles, and obtained exactly $0$ everywhere, in turn also compatible with the structure of a non-chiral quantum spin liquid. 

\begin{table}[ht]
	\centering
	\begin{tabular}{||c|c|c||}
	\hline 
		Model & ~No symmetry ~& ~$SU(2)$ ~ \\
		\hline
		$s=1$ BLBQ & $(7, 0.3188)$ & $(6, 19.5, 0.3108)$ \\ 
		$s =1/2$ KHAF & $(10, -0.4348)$ & $(7, 17.75, -0.4349)$ \\ 
		$s=2$ KHAF & $(10, -4.7975)$ & $(5, 19, -4.8227)$ \\ 
		\hline
    \end{tabular}
    \caption{Ground state energies obtained for the maximum achievable bond dimension for the bilinear-biquadratic (BLBQ) and KHAF models that we considered. We show $(D, e_0)$ for non-symmetric simulations and $(D_\text{sym}, D_\text{eff}, e_0)$ for $SU(2)$-invariant ones, with $D$ the non-symmetric bond dimension, $D_\text{sym}$ the symmetric bond dimension and $D_\text{eff}$ the effective bond dimension when using $SU(2)$ averaged for all bonds, which can be integer or fractional.} 
    \label{tab0} 
\end{table}

\vspace{5pt}

{\color{blue} {\it Conclusions.-}} In Table~\ref{tab0} we make a comparison of the computed ground state energies for the maximum achievable bond dimensions, for the three models considered here, and for non-symmetric and $SU(2)$-symmetric simulations. We conclude that implementing $SU(2)$ symmetry in 2d TN algorithms usually produces better energies than non-symmetric simulations, but the performance depends on the specifics of the model and in particular on the gap of the phase being targeted. For the spin-$2$ model, the $SU(2)$ simulations point towards a quantum spin liquid as a plausible ground state. 

{\bf Acknowledgments.-} We acknowledge Andreas Haller, Saeed Jahromi, Matteo Rizzi, and Sukhbinder Singh for insightful discussions. We also acknowledge DFG funding through project GZ OR 381/3-1, as well as the MAINZ Graduate School of Excellence.

\end{document}